\documentclass[12pt,draftcls,onecolumn]{IEEEtran}
\usepackage{amsmath,amsfonts,amsthm,amsxtra} %amsthm gives dots after theorem numbers
\usepackage[cp866]{inputenc}
\usepackage[english]{babel}
\usepackage{array}
\usepackage{enumitem}
\usepackage{hyperref}
\usepackage{bm}                                     %+ bold math
\usepackage{epsfig}
\usepackage[usenames,dvipsnames]{pstricks}
\usepackage{pst-node} % For axes

\def\blsm{1.1}

\def\ph{\varphi}
\def\baselinestretch{\blsm} %!!
\def\rank{\operatorname{rank}}                   %+
\def\tr{\operatorname{tr}}                       %+
\def\diag{\operatorname{diag}}                   %+
\def\ind{\operatorname{ind}}                     %+
\def\C{\mathbb{C}}                               %+
\def\NN{\mathop{\mathcal N}\nolimits}            %+
\def\RR{\mathop{\mathcal R}\nolimits}            %+

\def\sqa{\sqcap\!\!\!\!\sqcup}
\def\epr{\hfill$\sqa$\smallskip}                 %+
\def\squareforqed{\hbox{\rlap{$\sqcap$}$\sqcup$}}
\def\qed{\ifmmode\squareforqed\else{\unskip\nobreak\hfil\penalty50\hskip1em\null\nobreak\hfil\squareforqed
                                                         \parfillskip=0pt\finalhyphendemerits=0\endgraf}\fi}
\def\liml {\mathop{\lim}  \limits}               %+
\def\Re {\mathop{{\rm Re}}\nolimits}             %+
\def\aa{\alpha}                                  %+

\def\G{\Gamma}                                   %+
\def\la{\lambda}                                 %+
\def\cdc{,\ldots,}                               %+
\def\1n{1,\ldots,n}                              %+
\def\J {\mathop{\bar{J}}\nolimits}               %+
\def\vj{\mathop{\tilde{J}}\nolimits}             %+

\def\xz{\hspace{-.07em}}
\def\xy{\hspace{.07em}}
\accentedsymbol{\Pin}{\stackrel{\;\scriptscriptstyle\infty}{P_{\mathstrut}}} %+
\def\beq#1{\begin{equation}\label{#1}}
\def\eeq{\end{equation}}
\def\beqq{\begin{eqnarray}}
\def\eeqq{\end{eqnarray}}
\def\beg{\begin{gather*}}

\newcommand{\R}{\mathbb{R}}

\def\iteme{\em\item\em}
\def\iI{\hspace*{5.3ex}}

\newtheorem{theorem}{Theorem}{\bfseries}{\itshape}
\newtheorem{proposition}{Proposition}{\bfseries}{\itshape}
\newtheorem{corol}{Corollary}{\bfseries}{\itshape}
\newtheorem{defn}{Definition}{\bfseries}{\bfseries}
\begin{document}
\title{The forest consensus theorem}

\author{Pavel~Chebotarev,        Rafig~Agaev
\thanks{P.~Chebotarev and R.~Agaev are with the Institute of Control Sciences of the Russian Academy of Sciences, 65 Profsoyuznaya Street, Moscow 117997, Russia.}}

% The paper headers
\markboth{Journal of \LaTeX\ Class Files,~Vol.~XX, No.~X, ~2013}%
{Shell \MakeLowercase{\textit{et al.}}: The forest consensus theorem}
\maketitle

% As a general rule, do not put math, special symbols or citations
% in the abstract or keywords.
\begin{abstract}
We show that the limiting state vector in the differential model of consensus seeking with an arbitrary communication digraph is obtained by multiplying the eigenprojection of the Laplacian matrix of the model by the vector of initial states. Furthermore, the eigenprojection coincides with the stochastic matrix of maximum out-forests of the weighted communication digraph. These statements make the forest consensus theorem. A~similar\- result for DeGroot's iterative pooling model requires the Ces\`aro (time-average) limit in the general case.
The forest consensus theorem is useful for the analysis of consensus protocols.
\end{abstract}

% Note that keywords are not normally used for peerreview papers.

% For peer review papers, you can put extra information on the cover
% page as needed:
% \ifCLASSOPTIONpeerreview
% \begin{center} \bfseries EDICS Category: 3-BBND \end{center}
% \fi
%
% For peerreview papers, this IEEEtran command inserts a page break and
% creates the second title. It will be ignored for other modes.
\IEEEpeerreviewmaketitle

\section{Introduction}
\label{s_intro}

\IEEEPARstart{T}he continuous-time model of consensus seeking in a multiagent system has the form \cite{Olfati-SaberMurray04,CheAga09ARC} %?% remove our
\beqq
\label{e_model1}%+
\dot x_i(t)&=&u_i(t),\\
     u_i(t)&=&-\sum_{j=1}^na_{ij}\left({x_i(t)-x_j(t)}\right),\quad i=\1n,
\label{e_model2}%+
\eeqq
where $x_i(t)$ is the state of the $i$'th agent and $a_{ij}\ge0$ is the weight with which agent $i$ takes into account the discrepancy with agent~$j.$ The matrix form of the model \eqref{e_model1}--\eqref{e_model2} is:
\beq{e_modeL}%+
\dot x(t)=-L\,x(t),
\eeq
where $x(t)=(x_1(t)\cdc x_n(t))^T,$ $L$ is the \emph{Laplacian matrix of the model\/} \eqref{e_model1}--\eqref{e_model2}:
\beq{e_L}%+
L=\diag(A\xy\bm1)-A,
\eeq
$A=(a_{ij})_{n\times n},\,$ and $\bm1=(1\cdc 1)^T.$

The nonsymmetric Laplacian matrices of this kind were studied in \cite{AgaChe00,CheAga02ap,AgaChe05LAA}.

In this paper, we present the \emph{forest consensus theorem\/} stating that for an arbitrary non-negative matrix $A$ and any trajectory $x(t)$ satisfying \eqref{e_model1}--\eqref{e_model2},
\[%\beq{e_AsympState0}%-
\lim_{t\to\infty}x(t)=\vj x(0)
\]%\eeq
holds, where $\vj$ is %\footnote{The special case where $A$ is diagonalizable and has rank $n-1$ was considered in~\cite{Aga12UBSe}.}
the eigenprojection of~$L,$ which coincides with the matrix $\J$ of maximum out-forests of the communication digraph corresponding to~$A.$

A similar result, which involves the Ces\`aro limit, holds for the discretization of the model \eqref{e_model1}--\eqref{e_model2}.

The paper is organized as follows. After introducing the necessary notation and summarizing the preliminary results, in
Section\;\ref{s_main} we prove the forest consensus theorem.
Section\;\ref{s_property} is devoted to the properties of the limiting state of the model.
Section\;\ref{s_example} contains a numerical example; in
Section\;\ref{s_d=1}, we show that the classical results on a communication digraph having a spanning diverging tree immediately follow from the forest consensus theorem. Finally, Section\;\ref{s_DeGroot} presents a counterpart of the forest consensus theorem for the discretization of the model \eqref{e_model1}--\eqref{e_model2}.

\section{Basic concepts and preliminary results}
\label{s_notat}

\subsection{Eigenprojections and functions of matrices}
\label{s_eigenpr}

Let $A\in\C^{n\times n}$ be an arbitrary square matrix.
Let $\nu=\ind A$ be the {\em index\/} of $A,$ i.e., the smallest ${k}\in\{0,1,\ldots\}$ such that $\rank A^{k+1}=\rank A^k,$ where $A^0$ is identified with the identity matrix~$I.$ $A$~is nonsingular iff $\nu=0.$ The index of a singular matrix is the index of its eigenvalue $0,$ i.e., the multiplicity of $0$ as the root of its minimal polynomial, or, equivalently, the size of the largest Jordan block with zero diagonal in its Jordan form.  If $\nu=1$ then the algebraic and geometric multiplicities of $0$ coincide (in this case, the eigenvalue $0$ of $A$ is called \emph{semisimple}). %($0,$ if there are no such blocks).

Let $\RR(A)$ and $\NN(A)$ be the range and the null space of~$A,$ respectively. The {\em eigenprojection} \cite{Rothblum76ai} of $A$ {\em at eigenvalue\/}~$0$ is\footnote{Such an eigenprojection is also called the {\em principal idempotent\/} \cite{Hartwig76}.} a projection (i.e.\ an idempotent matrix) $Z$ such that $\RR(Z)=\NN(A^{\nu})$ and ${\NN}(Z)=\RR(A^{\nu}).$ In other words, $Z$ is a projection {\em to $\NN(A^{\nu})$ along $\RR(A^{\nu}).$} In the case of a singular matrix $A,$ following \cite{Rothblum76}, we call $Z$ the \emph{eigenprojection of~$A$\/} (without mentioning eigenvalue~$0$). The eigenprojection is unique because an idempotent matrix is uniquely determined by its range and null space (see, e.g., \cite[Sections\;2.4 and\;2.6]{Ben-IsraelGreville03}).

Eigenprojections underlie the definition of the \emph{components\/} of a matrix which, in turn, are used to define $\ph(A)$ for differentiable functions $\ph:\C\to\C$ (see either of \cite[Chapter\:5]{Gantmacher60}, \cite[Section\;2.5]{Ben-IsraelGreville03}, \cite{LancasterTismenetsky85,Higham08}), in the theory of generalized inverse matrices, as well as in the numerous applications of matrix analysis.

Let $\la_1\cdc\la_s$ be all \emph{distinct\/} eigenvalues of~$A;$ let $\nu_k$ be the \emph{index of\/} $\la_k$ defined as the index of $A-\la_kI.$
According to the theory of matrix components \cite[Chapter\:5]{Gantmacher60}, for any function $\ph:\C\to\C$ having finite derivatives $\ph^{(j)}(\la_k)$ of the first $\nu_k-1$ orders at $\la_1\cdc\la_s$, $\ph(A)$ is defined as follows:
\beq{e_funcmatr}%+
\ph(A):=\sum_{k=1}^s\sum_{j=0}^{\nu_k-1}\ph^{(j)}(\la_k)\,Z_{kj},
 \eeq
where the derivative $\ph^{(0)}$ of order $0$ is the value of $\ph$ and $Z_{kj}$ are the {\em components of}~$A$ defined by
\beq{e_comp}
Z_{kj}=(j!)^{-1}(A-\la_kI)^j\,Z_{k0}.
\eeq
Here, the component $Z_{k0}$ is the eigenprojection of $A-\la_kI$ $(k=1\cdc s)$ also called the \emph{eigenprojection of $A$ at~$\la_k.$}

For more details on eigenprojections, see, e.g., \cite{AgaChe02}.

\subsection{The stochastic matrix of maximum out-forests}
\label{s_forests}

A matrix $A=(a_{ij})$ of the model \eqref{e_model1}--\eqref{e_model2} determines a weighted \emph{communication digraph\/} $\G$ with vertex set $V(\G)=\{\1n\}$: $\G$ has the $(j,i)$ arc with weight $w_{\xz ji}=a_{ij}$ whenever $a_{ij}>0$ (i.e., when agent $j$ influences agent~$i$). Thus, arcs of $\G$ are oriented \emph{in the direction of influence}; the weight of an arc is the degree of influence.

A {\it diverging tree\/} is a weakly connected (i.e., its corresponding undirected graph is connected) digraph in which one vertex, called the {\it root}, has indegree zero and the other vertices have indegree one.
A diverging tree is said to {\em diverge\/} from its root. Spanning diverging trees are also called {\em out-arborescences\/} or {\em out-branchings\/} \cite{MesbahiEgerstedt10book}.
A~{\em diverging forest\/} is a digraph all of whose weak components (i.e., maximal weakly connected subdigraphs) are diverging trees. The roots of these trees form the \emph{set of roots\/} of the diverging forest.

\begin{defn}
{\rm Any spanning diverging forest of a digraph $\G$ is called an {\em out-forest\/} of~$\G.$
An out-forest $F$ of $\G$ is a {\em maximum out-forest\/} of $\G$ if $\G$ has no out-forest with a greater number of arcs than in~$F$.
The \emph{out-forest dimension of\/} $\G$ is the number of components in any maximum out-forest of~$\G.$
}
\end{defn}

The \emph{weight\/} of a weighted digraph is the product of its arc weights. The \emph{matrix $\J=(\J_{ij})$ of maximum out-forests\/} of a weighted digraph $\G$ is defined as follows:
\beq{e_elJ}
\J_{ij}=\frac{f_{ij}}f,\quad i,j=\1n,
\eeq
where $f$ is the total weight of all maximum out-forests of~$\G$ and $f_{ij}$ is the total weight of those of them that have $i$ belonging to a tree diverging from~$j.$
In Proposition\:\ref{p_eigproj}, we list some properties of $L$ and~$\J$ (cf.\ \cite{CheAga09ARC,Cheb10PIEEE,AgaChe11ARCE1}) which are useful for the analysis of consensus protocols.

\begin{proposition}\label{p_eigproj}
Let $L$ be the Laplacian matrix of the model \eqref{e_model1}--\eqref{e_model2}.
Let $\J$ be the matrix of maximum out-forests of the corresponding communication digraph\/~$\G$ whose out-forest dimension is~$d.$ Then\/$:$
 \begin{enumerate}[noitemsep,nolistsep]%[label=\roman(*)]
   \iteme\label{i_si}  $L$ is singular $($since $L\bm1=0);$
   \iteme\label{i_po}  If\/ $\la\ne0$ is an eigenvalue of\/ $L,$ then $\Re(\la)>0\;\emph{\cite[\emph{Proposition}\:9]{AgaChe01}};$
   \iteme\label{i_ind} $\ind L=1\:\;\emph{\cite[\emph{Proposition}\:12]{CheAga02ap}};$
   \iteme\label{i_ra}  $\rank L=n-d;\,$ $\rank\J=\tr\J=d\:\;\emph{\cite[\emph{Proposition}\:11]{AgaChe00}};$
   \iteme\label{i_1}   $\J$ is row stochastic as by definition$,$ $\sum_{j=1}^nf_{ij}=f$ for any\/ $i\in\{\1n\};$
   \iteme\label{i_ep}  $\J$ is the eigenprojection of\/~$L\:\;\emph{\cite[\emph{Proposition}\:12]{CheAga02ap}},$ which implies that $\J^2=\J;$
   \iteme\label{i_0}   $L\!\J=\J\!L=0\:\;\emph{\cite[\emph{Theorem}\;5]{AgaChe00}};\,$ $\NN(\J)=\RR(L),\;\RR(\J)=\NN(L)$ $($by items\;$\ref{i_ind}$ and\;$\ref{i_ep});$
   \iteme\label{i_li}  $\J=\liml_{{\aa}\to\infty}(I+\aa L)^{-1}\:\;\emph{\cite[\emph{Theorem}\;6]{AgaChe00}};$
%  \iteme\label{i_02}  $\J=p_q^{-1} h(L),$ where \,$\la h(\la)=\la(\la^q+p_1\la^{q-1}+\ldots+p_q)$ is the minimal polynomial of\;$L$ and\, $q=n-d \;\emph{\cite[\emph{Theorem}\:1]{AgaChe02}};$
   \iteme\label{i_Ga}  ${\J={C(0)\big/h(0)}}$, where\footnote{In some cases, the expression \cite[Theorem\:1]{AgaChe02} can be more convenient for calculations.} $C(\la)$ is the quotient of the matrix polynomial $\la h(\la)I$ and the binomial $\la I-L,\,$ $\la h(\la)$ being the minimal polynomial of\;$L$ $($this follows from \emph{\cite[\emph{Eq.}\,(22) \emph{in Chapter}\:5]{Gantmacher60}}$);$
   \iteme\label{i_01}  $\J=\J_{n-d},$ where $\J_{n-d}$ is defined recursively\/$:$
   $\J_k=I-k\frac{L\J_{k-1}}{\tr(L^{\mathstrut}\J_{k-1})},\,$ $k=1\cdc{n-d},$ $\J_0=I,$ and $L\J_{n-d}=0\,$ \emph{\cite[\emph{Section}\;4]{AgaChe01} \emph{or} \cite[\emph{Section}\;5]{CheAga02ap}}.
 \end{enumerate}
\end{proposition}
%   \item[\rm ()] $\J$ is idempotent\/$:$ $\J^2=\J;$

\smallskip
An elementwise characterization of $\J$ is given in \cite[Theorem\:$2'$]{AgaChe00}.

\section{The forest consensus theorem}
\label{s_main}

\begin{theorem}
\label{th_main}
Let $x(t)$ be a solution of \eqref{e_modeL}. Then
\beq{e_AsympState}%-
\lim_{t\to\infty}x(t)=\vj x(0),
\eeq
where $\vj$ is the eigenprojection of~$L.$ Moreover$,$ $\vj$ coincides with the matrix $\J$ of maximum out-forests of the communication digraph corresponding to~$L.$
\end{theorem}

\noindent
\textbf{Proof.} %of theorem \ref{th_main}.}
All solutions of \eqref{e_modeL} satisfy the identity \cite[Eq.\:(46) in Chapter\:5]{Gantmacher60}
\beq{e_SpecSolu}%+
x(t)=e^{-Lt}x(0).
\eeq

According to \eqref{e_funcmatr} $e^{-Lt}$ is representable in the form (cf.\ Eq.\,(12) of Chapter\:4 in~\cite{Gantmacher59-2})
\beq{e_matrExp}%+imp
e^{-Lt}=\sum_{k=1}^s \sum_{j=0}^{\nu_k-1}Z_{kj}\,t^{j}e^{-\la_k t},
\eeq
where $\la_1\cdc \la_s$ are all distinct eigenvalues of~$L.$

Since $L$ is singular, we can set $\la_1=0.$ Then $Z_{1j}$ are the components of $L$ corresponding to the characteristic root~$0$.
By item\:\ref{i_ind} of Proposition\:\ref{p_eigproj}, $\nu_1=\ind L=1$ and
by item\:\ref{i_ep}  of Proposition\:\ref{p_eigproj}, $Z_{10},$ the eigenprojection of $L$ denoted by $\vj,$ coincides with the matrix~$\J$ of maximum out-forests of the communication digraph~$\G.$

Since the components $Z_{kj}$ of $L$ are independent of $t$ while, by item\:\ref{i_po} of Proposition\:\ref{p_eigproj}, ${\Re(\la_k)>0}$ $(k\ge2),$ we have
\beq{e_SimpliComp}%+
\lim_{t \to\infty}\sum_{k=2}^s\sum_{j=0}^{\nu_k-1}Z_{kj}\,t^{j}e^{-\la_k t}=0.
\eeq

Finally, \eqref{e_SpecSolu}--\eqref{e_SimpliComp} and $\nu_1=1$ yield
$$
\lim_{t\to\infty}x(t)
=\lim_{t \to\infty} e^{-Lt}x(0)
=Z_{10}\,x(0)=\vj x(0).\eqno\sqa
$$
%Theorem\;\ref{th_main} is proved.\epr

%The special case of Theorem\:\ref{th_main} in which $A$ is diagonalizable and has rank $n-1$ was considered in~\cite{Aga12UBSe}.

\section{The properties of the asymptotic state}
\label{s_property}

Now we need some additional notation.
A \emph{basic bicomponent\/} of a digraph $\G$ is any maximal (by inclusion) strongly connected weighted subdigraph of $\G$ such that there are no arcs coming into it from outside.
By \cite[Proposition\:6]{AgaChe00}, the number of basic bicomponents in $\G$ is equal to the out-forest dimension $d$ of~$\G.$

Let $x(\infty)$ be the limiting state vector of the model \eqref{e_model1}--\eqref{e_model2}: $x(\infty)=\lim_{t\to\infty}x(t).$

\begin{corol}[of Theorem\:\ref{th_main}]\label{c_prop}
Let $K$ be a basic bicomponents of\/ $\G;$ let $j$ be a vertex of $K.$ It holds that\/$:$
\begin{enumerate}[noitemsep,nolistsep]%[label=\roman(*)]
   \iteme\label{i_ij1} If $i$ is a vertex of $K$ or $i$ is reachable $($by a directed path$)$ from $K$ and not reachable from the other basic bicomponents of\/ $\G,$ then $x_i(\infty)=x_j(\infty)$ and $x_i(\infty)$ is equal to the value of consensus for the communication digraph $K$ alone\xy$;$
   \iteme\label{i_ij2} If vertex $i$ is reachable from several basic bicomponents of\/ $\G,$ then $x_i(\infty)$ is between the minimum and maximum elements of $x(\infty)$ that correspond to these basic bicomponents $($and is strictly between them if the minimum and maximum differ$);$
   \iteme\label{i_ij3} If vertex $i$ is not in a basic bicomponent of\/ $\G,$ then $x(\infty)$ is independent of\/ $x_i(0).$
\end{enumerate}
\end{corol}

Corollary\:\ref{c_prop} is easily proved using the row stochasticity of $\J$ and two simple facts which follow from \cite[Theorem\:2$'$]{AgaChe00}. The facts are: (1)~$\J_{ij}\ne0$ if and only if $j$ belongs to a basic bicomponent of $\G$ and $i$ is reachable from~$j;$ (2)~If $i$ and $j$ belong to the same basic bicomponents in $\G,$ then the $i$-row and $j$-row of $\J$ are equal, while the $i$-column and $j$-column are proportional.

Using time shift and item\:\ref{i_0} of Proposition\;\ref{p_eigproj} we have\footnote{Cf.\ \cite{AmelinyGranichiny12}, equation between (16) and (17).}
\begin{corol}[of Theorem\:\ref{th_main}]\label{c_shift}
Let $x(t)$ be a solution of \eqref{e_modeL}. Then for any $t\in\R,\,$ $\J x(t)=x(\infty).$ Consequently$,$ for any $t_1,t_2\in\R,\,$ $\J(x(t_1)-x(t_2))=0,$ i.e.$,$ $(x(t_1)-x(t_2))\in\NN(\J)=\RR(L).$
\end{corol}

\section{Example}
\label{s_example}

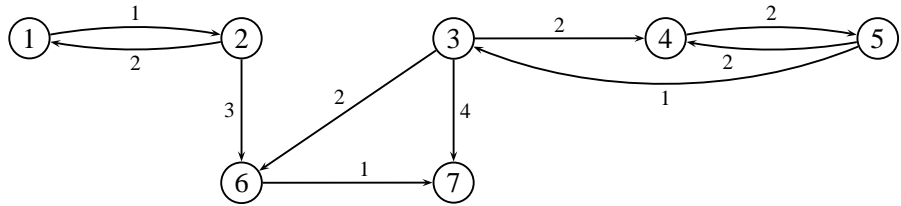
\begin{figure}[h*]
%\setlength{\unitlength}{1.3mm}
%\scalebox{0.9} % Change this value to rescale the drawing.
\psscalebox{0.9 0.9}{
\begin{pspicture}(-3,0)(3,3.0)
 \psmatrix[mnode=circle,colsep=2.5] 1 & 2 & 3 & 4 & 5 \\ & 6 & 7 \endpsmatrix
 \psset{shortput=nab,arrows=->,labelsep=2pt}
 {\footnotesize
 \ncarc[arcangle=10]{1,1}{1,2}^{1}
 \ncarc[arcangle=10]{1,2}{1,1}^{2}
 \ncline{1,2}{2,2}_{3}
 \ncline{2,2}{2,3}^[npos=.6]{1}
 \ncline{1,3}{2,2}_{2}
 \ncline{1,3}{2,3}^{4}
 \ncline{1,3}{1,4}^{2}
 \ncarc[arcangle=10]{1,4}{1,5}^{2}
 \ncarc[arcangle=10]{1,5}{1,4}^[npos=.75]{2}
 \ncarc[arcangle=22.5]{1,5}{1,3}^[npos=.5]{1}}
\end{pspicture}}
\caption{\label{fig_ex1}A communication digraph $\G.$}
\end{figure}

Consider the weighted communication digraph $\G$ shown in Fig.\,\ref{fig_ex1}. It has two basic bicomponents whose vertex sets are $\{1,2\}$ and $\{3,4,5\}.$

The Laplacian matrix \eqref{e_L} of the model \eqref{e_model1}--\eqref{e_model2} corresponding\footnote{On this correspondence, see Section\;\ref{s_forests}.} to $\G$ is:
{\small
\[L=\left(
  \begin{array}{rrrrrrr}
 2   &-2    &  0     &  0    &  0     &  0     & 0    \\
-1   & 1    &  0     &  0    &  0     &  0     & 0    \\
 0   & 0    &  1     &  0    & -1     &  0     & 0    \\
 0   & 0    & -2     &  4    & -2     &  0     & 0    \\
 0   & 0    &  0     & -2    &  2     &  0     & 0    \\
 0   &-3    & -2     &  0    &  0     &  5     & 0    \\
 0   & 0    & -4     &  0    &  0     & -1     & 5    \\
  \end{array}
\right).
\]
}

The spectrum of $L$ is real: $(0,\, 0,\, 2,\, 3,\, 5,\, 5,\, 5),$ which is not generally the case for a Laplacian matrix of a weighted digraph. On the other hand, $L$ is not diagonalizable as the geometric multiplicity of the triple eigenvalue $5$ is~$1.$
The minimal polynomial of~$L$ is
\[%\beq{e_PsiL}%-
%\psi_L(\la)
\la h(\la)=\la(\la-2)(\la-3)(\la-5)^3.
\]%\eeq

To find $Z_{10}=\vj,$ one can use item\:\ref{i_Ga} of Proposition\:\ref{p_eigproj}:
\beq{e_Z10}%+
Z_{10}=\frac{C(0)}{h(0)},
\eeq
where $h(0)=(-2)(-3)(-5)^3=-750,$ while $C(0)$ can be determined using (55) and (56) in \cite[Chapter\:4]{Gantmacher60}:
$\Psi(0,\mu)={(\mu-2)(\mu-3)(\mu-5)^3}$ and
\beq{e_C(0)}
C(0)=\Psi(0\cdot I,\,L)=(L-2\xy I)(L-3\xy I)(L-5\xy I)^3.
\eeq

Substituting $C(0)$ and $h(0)$ into \eqref{e_Z10} yields

\beq{e_Jex}
Z_{10}=\vj=\J=\frac1{750}\left(
\begin{array}
{rrrrrrr}
250& 500&   0&   0&   0& 0& 0\\
250& 500&   0&   0&   0& 0& 0\\
  0&   0& 300& 150& 300& 0& 0\\
  0&   0& 300& 150& 300& 0& 0\\
  0&   0& 300& 150& 300& 0& 0\\
150& 300& 120&  60& 120& 0& 0\\
 30&  60& 264& 132& 264& 0& 0\\
\end{array}
\right).
\eeq

The matrix $\J$ can also be found using item\:\ref{i_01} of Proposition\:\ref{p_eigproj}: as well as \eqref{e_C(0)}, this involves four matrix multiplications, but it does not require the knowledge of the nonzero eigenvalues or the minimal polynomial of~$L.$ Direct enumeration of forests has no practical value for the computation of~$\J.$ However, to give the reader a little taste of this ``forestry'', in Fig.\,2 we present all 32 maximum out-forests of $\G$ and their weights. The total weight of them is $f\!=\!750;$ by the definition \eqref{e_elJ}, it is the common denominator of the entries of~$\J=(\J_{ij}).$ The numerator of $\J_{ij}$ is the total weight $f_{ij}$ of the maximum out-forests in which $i$ belongs to a tree diverging from~$j.$ For example, for $\J_{65},$ these are the forests \#10, 12, 14, 16, 26, 28, 30, and 32 whose weights are 16, 4, 16, 4, 32, 8, 32, and 8, respectively, so that $f_{65}=120$ and $\J_{65}=\frac{120}{750},$ in concordance with~\eqref{e_Jex}.
\pagebreak

%ROW 1
\hspace*{-1.5em}\psscalebox{0.598 0.598}{\begin{pspicture}(0,0)(0.6,0)
\psmatrix[mnode=circle,colsep=0.892] 1 & 2 & 3 & 4 & 5 \\ & 6 & 7 \endpsmatrix
\psset{shortput=nab,arrows=->,labelsep=2pt} {\footnotesize \ncarc[arcangle=10]{1,1}{1,2}^{1}                                   \ncline{1,2}{2,2}_{3} \ncline{2,2}{2,3}^{1}
                                            \ncline{1,3}{1,4}^{2} \ncarc[arcangle=10]{1,4}{1,5}^{2}                                                                              }\\
\end{pspicture}\begin{pspicture}(0,0)(0.6,3.5)
\psmatrix[mnode=circle,colsep=0.892] 1 & 2 & 3 & 4 & 5 \\ & 6 & 7 \endpsmatrix
\psset{shortput=nab,arrows=->,labelsep=2pt} {\footnotesize \ncarc[arcangle=10]{1,1}{1,2}^{1}
\ncline{1,3}{2,2}_{2} \ncline{1,3}{2,3}^{4} \ncline{1,3}{1,4}^{2} \ncarc[arcangle=10]{1,4}{1,5}^{2}                                                                              }\\
\end{pspicture}\begin{pspicture}(0,0)(0.6,3.5)
\psmatrix[mnode=circle,colsep=0.892] 1 & 2 & 3 & 4 & 5 \\ & 6 & 7 \endpsmatrix
\psset{shortput=nab,arrows=->,labelsep=2pt} {\footnotesize \ncarc[arcangle=10]{1,1}{1,2}^{1}                                   \ncline{1,2}{2,2}_{3}
                      \ncline{1,3}{2,3}^{4} \ncline{1,3}{1,4}^{2} \ncarc[arcangle=10]{1,4}{1,5}^{2}                                                                            }\\
\end{pspicture}\begin{pspicture}(0,0)(0.6,3.5)
\psmatrix[mnode=circle,colsep=0.892] 1 & 2 & 3 & 4 & 5 \\ & 6 & 7 \endpsmatrix
\psset{shortput=nab,arrows=->,labelsep=2pt} {\footnotesize \ncarc[arcangle=10]{1,1}{1,2}^{1}                                                         \ncline{2,2}{2,3}^{1}
\ncline{1,3}{2,2}_{2}                       \ncline{1,3}{1,4}^{2} \ncarc[arcangle=10]{1,4}{1,5}^{2}                                                                            }\\
\end{pspicture}
}
\vspace*{-1em}{\footnotesize\;\ Forest \#1; weight$\;=\!12$\hspace*{1.9em}\; {\iI}Forest \#2; weight$\;=\!32$\hspace*{1.9em}\, {\iI}Forest \#3; weight$\;=\!48$\hspace*{1.9em}\;\, {\iI}Forest \#4; weight$\;=\!8$}

%ROW 2
\hspace*{-1.5em}\psscalebox{0.598 0.598}{\begin{pspicture}(0,0)(0.6,3.8)
\psmatrix[mnode=circle,colsep=0.892] 1 & 2 & 3 & 4 & 5 \\ & 6 & 7 \endpsmatrix
\psset{shortput=nab,arrows=->,labelsep=2pt} {\footnotesize \ncarc[arcangle=10]{1,1}{1,2}^{1}                                   \ncline{1,2}{2,2}_{3} \ncline{2,2}{2,3}^{1}
                                                                  \ncarc[arcangle=10]{1,4}{1,5}^{2}                                             \ncarc[arcangle=36]{1,5}{1,3}^{1}}\\
\end{pspicture}\begin{pspicture}(0,0)(0.6,2.5)
\psmatrix[mnode=circle,colsep=0.892] 1 & 2 & 3 & 4 & 5 \\ & 6 & 7 \endpsmatrix
\psset{shortput=nab,arrows=->,labelsep=2pt} {\footnotesize \ncarc[arcangle=10]{1,1}{1,2}^{1}
\ncline{1,3}{2,2}_{2} \ncline{1,3}{2,3}^{4}                       \ncarc[arcangle=10]{1,4}{1,5}^{2}                                             \ncarc[arcangle=36]{1,5}{1,3}^{1}}\\
\end{pspicture}\begin{pspicture}(0,0)(0.6,2.5)
\psmatrix[mnode=circle,colsep=0.892] 1 & 2 & 3 & 4 & 5 \\ & 6 & 7 \endpsmatrix
\psset{shortput=nab,arrows=->,labelsep=2pt} {\footnotesize \ncarc[arcangle=10]{1,1}{1,2}^{1}                                   \ncline{1,2}{2,2}_{3}
                      \ncline{1,3}{2,3}^{4}                       \ncarc[arcangle=10]{1,4}{1,5}^{2}                                             \ncarc[arcangle=36]{1,5}{1,3}^{1}}\\
\end{pspicture}\begin{pspicture}(0,0)(0.6,3.8)
\psmatrix[mnode=circle,colsep=0.892] 1 & 2 & 3 & 4 & 5 \\ & 6 & 7 \endpsmatrix
\psset{shortput=nab,arrows=->,labelsep=2pt} {\footnotesize \ncarc[arcangle=10]{1,1}{1,2}^{1}                                                         \ncline{2,2}{2,3}^{1}
\ncline{1,3}{2,2}_{2}                                             \ncarc[arcangle=10]{1,4}{1,5}^{2}                                             \ncarc[arcangle=36]{1,5}{1,3}^{1}}\\
\end{pspicture}}
\vspace*{-1em}{\footnotesize\;\;\,\ Forest \#5; weight$\;=\!6$\hspace*{1.9em}\;\, {\iI}Forest \#6; weight$\;=\!16$\hspace*{1.9em}\; {\iI}Forest \#7; weight$\;=\!24$\hspace*{1.9em}\;\, {\iI}Forest \#8; weight$\;=\!4$}

%ROW 3
\hspace*{-1.5em}\psscalebox{0.598 0.598}{\begin{pspicture}(0,0)(0.6,3.8)
\psmatrix[mnode=circle,colsep=0.892] 1 & 2 & 3 & 4 & 5 \\ & 6 & 7 \endpsmatrix
\psset{shortput=nab,arrows=->,labelsep=2pt} {\footnotesize \ncarc[arcangle=10]{1,1}{1,2}^{1}                                   \ncline{1,2}{2,2}_{3} \ncline{2,2}{2,3}^{1}
                                                                                                    \ncarc[arcangle=10]{1,5}{1,4}^[npos=.75]{2} \ncarc[arcangle=36]{1,5}{1,3}^{1}}\\
\end{pspicture}\begin{pspicture}(0,0)(0.6,4.0)
\psmatrix[mnode=circle,colsep=0.892] 1 & 2 & 3 & 4 & 5 \\ & 6 & 7 \endpsmatrix
\psset{shortput=nab,arrows=->,labelsep=2pt} {\footnotesize \ncarc[arcangle=10]{1,1}{1,2}^{1}
\ncline{1,3}{2,2}_{2} \ncline{1,3}{2,3}^{4}                                                         \ncarc[arcangle=10]{1,5}{1,4}^[npos=.75]{2} \ncarc[arcangle=36]{1,5}{1,3}^{1}}\\
\end{pspicture}\begin{pspicture}(0,0)(0.6,4.0)
\psmatrix[mnode=circle,colsep=0.892] 1 & 2 & 3 & 4 & 5 \\ & 6 & 7 \endpsmatrix
\psset{shortput=nab,arrows=->,labelsep=2pt} {\footnotesize \ncarc[arcangle=10]{1,1}{1,2}^{1}                                   \ncline{1,2}{2,2}_{3}
                      \ncline{1,3}{2,3}^{4}                                                         \ncarc[arcangle=10]{1,5}{1,4}^[npos=.75]{2} \ncarc[arcangle=36]{1,5}{1,3}^{1}}\\
\end{pspicture}\begin{pspicture}(0,0)(0.6,3.8)
\psmatrix[mnode=circle,colsep=0.892] 1 & 2 & 3 & 4 & 5 \\ & 6 & 7 \endpsmatrix
\psset{shortput=nab,arrows=->,labelsep=2pt} {\footnotesize \ncarc[arcangle=10]{1,1}{1,2}^{1}                                                         \ncline{2,2}{2,3}^{1}
\ncline{1,3}{2,2}_{2}                                                                               \ncarc[arcangle=10]{1,5}{1,4}^[npos=.75]{2} \ncarc[arcangle=36]{1,5}{1,3}^{1}}\\
\end{pspicture}}
\vspace*{-1em}{\footnotesize\;\;\,\ Forest \#9; weight$\;=\!6$\hspace*{1.9em}\, {\iI}Forest \#10; weight$\;=\!16$\hspace*{1.9em}\!\! {\iI}Forest \#11; weight$\;=\!24$\hspace*{1.9em}\! {\iI}Forest \#12; weight$\;=\!4$}

%ROW 4
\hspace*{-1.5em}\psscalebox{0.598 0.598}{\begin{pspicture}(0,0)(0.6,3.8)
\psmatrix[mnode=circle,colsep=0.892] 1 & 2 & 3 & 4 & 5 \\ & 6 & 7 \endpsmatrix
\psset{shortput=nab,arrows=->,labelsep=2pt} {\footnotesize \ncarc[arcangle=10]{1,1}{1,2}^{1}                                   \ncline{1,2}{2,2}_{3} \ncline{2,2}{2,3}^{1}
                                            \ncline{1,3}{1,4}^{2}                                                                               \ncarc[arcangle=36]{1,5}{1,3}^{1}}\\
\end{pspicture}\begin{pspicture}(0,0)(0.6,4.0)
\psmatrix[mnode=circle,colsep=0.892] 1 & 2 & 3 & 4 & 5 \\ & 6 & 7 \endpsmatrix
\psset{shortput=nab,arrows=->,labelsep=2pt} {\footnotesize \ncarc[arcangle=10]{1,1}{1,2}^{1}
\ncline{1,3}{2,2}_{2} \ncline{1,3}{2,3}^{4} \ncline{1,3}{1,4}^{2}                                                                               \ncarc[arcangle=36]{1,5}{1,3}^{1}}\\
\end{pspicture}\begin{pspicture}(0,0)(0.6,4.0)
\psmatrix[mnode=circle,colsep=0.892] 1 & 2 & 3 & 4 & 5 \\ & 6 & 7 \endpsmatrix
\psset{shortput=nab,arrows=->,labelsep=2pt} {\footnotesize \ncarc[arcangle=10]{1,1}{1,2}^{1}                                   \ncline{1,2}{2,2}_{3}
                      \ncline{1,3}{2,3}^{4} \ncline{1,3}{1,4}^{2}                                                                               \ncarc[arcangle=36]{1,5}{1,3}^{1}}\\
\end{pspicture}\begin{pspicture}(0,0)(0.6,3.8)
\psmatrix[mnode=circle,colsep=0.892] 1 & 2 & 3 & 4 & 5 \\ & 6 & 7 \endpsmatrix
\psset{shortput=nab,arrows=->,labelsep=2pt} {\footnotesize \ncarc[arcangle=10]{1,1}{1,2}^{1}                                                         \ncline{2,2}{2,3}^{1}
\ncline{1,3}{2,2}_{2}                       \ncline{1,3}{1,4}^{2}                                                                               \ncarc[arcangle=36]{1,5}{1,3}^{1}}\\
\end{pspicture}}
\vspace*{-1em}{\footnotesize\;\ Forest \#13; weight$\;=\!6$\hspace*{1.9em} {\iI}Forest \#14; weight$\;=\!16$\hspace*{1.9em}\! {\iI}Forest \#15; weight$\;=\!24$\hspace*{1.9em}\! {\iI}Forest \#16; weight$\;=\!4$}

%ROW 5
\hspace*{-1.5em}\psscalebox{0.598 0.598}{\begin{pspicture}(0,0)(0.6,3.8)
\psmatrix[mnode=circle,colsep=0.892] 1 & 2 & 3 & 4 & 5 \\ & 6 & 7 \endpsmatrix
\psset{shortput=nab,arrows=->,labelsep=2pt} {\footnotesize                                   \ncarc[arcangle=10]{1,2}{1,1}^{2} \ncline{1,2}{2,2}_{3} \ncline{2,2}{2,3}^{1}
                                            \ncline{1,3}{1,4}^{2} \ncarc[arcangle=10]{1,4}{1,5}^{2}                                                                             }\\
\end{pspicture}\begin{pspicture}(0,0)(0.6,4.0)
\psmatrix[mnode=circle,colsep=0.892] 1 & 2 & 3 & 4 & 5 \\ & 6 & 7 \endpsmatrix
\psset{shortput=nab,arrows=->,labelsep=2pt} {\footnotesize                                   \ncarc[arcangle=10]{1,2}{1,1}^{2}
\ncline{1,3}{2,2}_{2} \ncline{1,3}{2,3}^{4} \ncline{1,3}{1,4}^{2} \ncarc[arcangle=10]{1,4}{1,5}^{2}                                                                              }\\
\end{pspicture}\begin{pspicture}(0,0)(0.6,4.0)
\psmatrix[mnode=circle,colsep=0.892] 1 & 2 & 3 & 4 & 5 \\ & 6 & 7 \endpsmatrix
\psset{shortput=nab,arrows=->,labelsep=2pt} {\footnotesize                                   \ncarc[arcangle=10]{1,2}{1,1}^{2} \ncline{1,2}{2,2}_{3}
                      \ncline{1,3}{2,3}^{4} \ncline{1,3}{1,4}^{2} \ncarc[arcangle=10]{1,4}{1,5}^{2}                                                                              }\\
\end{pspicture}\begin{pspicture}(0,0)(0.6,3.8)
\psmatrix[mnode=circle,colsep=0.892] 1 & 2 & 3 & 4 & 5 \\ & 6 & 7 \endpsmatrix
\psset{shortput=nab,arrows=->,labelsep=2pt} {\footnotesize                                   \ncarc[arcangle=10]{1,2}{1,1}^{2}                       \ncline{2,2}{2,3}^{1}
\ncline{1,3}{2,2}_{2}                       \ncline{1,3}{1,4}^{2} \ncarc[arcangle=10]{1,4}{1,5}^{2}                                                                             }\\
\end{pspicture}}
\vspace*{-1em}{\footnotesize\,\ Forest \#17; weight$\;=\!24$\hspace*{1.9em}\!\! {\iI}Forest \#18; weight$\;=\!64$\hspace*{1.9em}\!\! {\iI}Forest \#19; weight$\;=\!96$\hspace*{1.9em}\!\!\! {\iI}Forest \#20; weight$\;=\!16$}

%ROW 6
\hspace*{-1.5em}\psscalebox{0.598 0.598}{\begin{pspicture}(0,0)(0.6,3.8)
\psmatrix[mnode=circle,colsep=0.892] 1 & 2 & 3 & 4 & 5 \\ & 6 & 7 \endpsmatrix
\psset{shortput=nab,arrows=->,labelsep=2pt} {\footnotesize                                   \ncarc[arcangle=10]{1,2}{1,1}^{2} \ncline{1,2}{2,2}_{3} \ncline{2,2}{2,3}^{1}
                                                                  \ncarc[arcangle=10]{1,4}{1,5}^{2}                                             \ncarc[arcangle=36]{1,5}{1,3}^{1}}\\
\end{pspicture}\begin{pspicture}(0,0)(0.6,4.0)
\psmatrix[mnode=circle,colsep=0.892] 1 & 2 & 3 & 4 & 5 \\ & 6 & 7 \endpsmatrix
\psset{shortput=nab,arrows=->,labelsep=2pt} {\footnotesize                                   \ncarc[arcangle=10]{1,2}{1,1}^{2}
\ncline{1,3}{2,2}_{2} \ncline{1,3}{2,3}^{4}                       \ncarc[arcangle=10]{1,4}{1,5}^{2}                                             \ncarc[arcangle=36]{1,5}{1,3}^{1}}\\
\end{pspicture}\begin{pspicture}(0,0)(0.6,4.0)
\psmatrix[mnode=circle,colsep=0.892] 1 & 2 & 3 & 4 & 5 \\ & 6 & 7 \endpsmatrix
\psset{shortput=nab,arrows=->,labelsep=2pt} {\footnotesize                                   \ncarc[arcangle=10]{1,2}{1,1}^{2} \ncline{1,2}{2,2}_{3}
                      \ncline{1,3}{2,3}^{4}                       \ncarc[arcangle=10]{1,4}{1,5}^{2}                                             \ncarc[arcangle=36]{1,5}{1,3}^{1}}\\
\end{pspicture}\begin{pspicture}(0,0)(0.6,3.8)
\psmatrix[mnode=circle,colsep=0.892] 1 & 2 & 3 & 4 & 5 \\ & 6 & 7 \endpsmatrix
\psset{shortput=nab,arrows=->,labelsep=2pt} {\footnotesize                                   \ncarc[arcangle=10]{1,2}{1,1}^{2}                       \ncline{2,2}{2,3}^{1}
\ncline{1,3}{2,2}_{2}                                             \ncarc[arcangle=10]{1,4}{1,5}^{2}                                             \ncarc[arcangle=36]{1,5}{1,3}^{1}}\\
\end{pspicture}}
\vspace*{-1em}{\footnotesize\,\ Forest \#21; weight$\;=\!12$\hspace*{1.9em}\!\! {\iI}Forest \#22; weight$\;=\!32$\hspace*{1.9em}\!\! {\iI}Forest \#23; weight$\;=\!48$\hspace*{1.9em}\! {\iI}Forest \#24; weight$\;=\!8$}

%ROW 7
\hspace*{-1.5em}\psscalebox{0.598 0.598}{\begin{pspicture}(0,0)(0.6,3.8)
\psmatrix[mnode=circle,colsep=0.892] 1 & 2 & 3 & 4 & 5 \\ & 6 & 7 \endpsmatrix
\psset{shortput=nab,arrows=->,labelsep=2pt} {\footnotesize                                   \ncarc[arcangle=10]{1,2}{1,1}^{2} \ncline{1,2}{2,2}_{3} \ncline{2,2}{2,3}^{1}
                                                                                                    \ncarc[arcangle=10]{1,5}{1,4}^[npos=.75]{2} \ncarc[arcangle=36]{1,5}{1,3}^{1}}\\
\end{pspicture}\begin{pspicture}(0,0)(0.6,4.0)
\psmatrix[mnode=circle,colsep=0.892] 1 & 2 & 3 & 4 & 5 \\ & 6 & 7 \endpsmatrix
\psset{shortput=nab,arrows=->,labelsep=2pt} {\footnotesize                                   \ncarc[arcangle=10]{1,2}{1,1}^{2}
\ncline{1,3}{2,2}_{2} \ncline{1,3}{2,3}^{4}                                                         \ncarc[arcangle=10]{1,5}{1,4}^[npos=.75]{2} \ncarc[arcangle=36]{1,5}{1,3}^{1}}\\
\end{pspicture}\begin{pspicture}(0,0)(0.6,4.0)
\psmatrix[mnode=circle,colsep=0.892] 1 & 2 & 3 & 4 & 5 \\ & 6 & 7 \endpsmatrix
\psset{shortput=nab,arrows=->,labelsep=2pt} {\footnotesize                                   \ncarc[arcangle=10]{1,2}{1,1}^{2} \ncline{1,2}{2,2}_{3}
                      \ncline{1,3}{2,3}^{4}                                                         \ncarc[arcangle=10]{1,5}{1,4}^[npos=.75]{2} \ncarc[arcangle=36]{1,5}{1,3}^{1}}\\
\end{pspicture}\begin{pspicture}(0,0)(0.6,3.8)
\psmatrix[mnode=circle,colsep=0.892] 1 & 2 & 3 & 4 & 5 \\ & 6 & 7 \endpsmatrix
\psset{shortput=nab,arrows=->,labelsep=2pt} {\footnotesize                                   \ncarc[arcangle=10]{1,2}{1,1}^{2}                       \ncline{2,2}{2,3}^{1}
\ncline{1,3}{2,2}_{2}                                                                               \ncarc[arcangle=10]{1,5}{1,4}^[npos=.75]{2} \ncarc[arcangle=36]{1,5}{1,3}^{1}}\\
\end{pspicture}}
\vspace*{-1em}{\footnotesize\,\ Forest \#25; weight$\;=\!12$\hspace*{1.9em}\!\! {\iI}Forest \#26; weight$\;=\!32$\hspace*{1.9em}\!\! {\iI}Forest \#27; weight$\;=\!48$\hspace*{1.9em}\! {\iI}Forest \#28; weight$\;=\!8$}

%ROW 8
\hspace*{-1.5em}\psscalebox{0.598 0.598}{\begin{pspicture}(0,0)(0.6,3.8)
\psmatrix[mnode=circle,colsep=0.892] 1 & 2 & 3 & 4 & 5 \\ & 6 & 7 \endpsmatrix
\psset{shortput=nab,arrows=->,labelsep=2pt} {\footnotesize                                   \ncarc[arcangle=10]{1,2}{1,1}^{2} \ncline{1,2}{2,2}_{3} \ncline{2,2}{2,3}^{1}
                                            \ncline{1,3}{1,4}^{2}                                                                               \ncarc[arcangle=36]{1,5}{1,3}^{1}}\\
\end{pspicture}\begin{pspicture}(0,0)(0.6,4.0)
\psmatrix[mnode=circle,colsep=0.892] 1 & 2 & 3 & 4 & 5 \\ & 6 & 7 \endpsmatrix
\psset{shortput=nab,arrows=->,labelsep=2pt} {\footnotesize                                   \ncarc[arcangle=10]{1,2}{1,1}^{2}
\ncline{1,3}{2,2}_{2} \ncline{1,3}{2,3}^{4} \ncline{1,3}{1,4}^{2}                                                                               \ncarc[arcangle=36]{1,5}{1,3}^{1}}\\
\end{pspicture}\begin{pspicture}(0,0)(0.6,4.0)
\psmatrix[mnode=circle,colsep=0.892] 1 & 2 & 3 & 4 & 5 \\ & 6 & 7 \endpsmatrix
\psset{shortput=nab,arrows=->,labelsep=2pt} {\footnotesize                                   \ncarc[arcangle=10]{1,2}{1,1}^{2} \ncline{1,2}{2,2}_{3}
                      \ncline{1,3}{2,3}^{4} \ncline{1,3}{1,4}^{2}                                                                               \ncarc[arcangle=36]{1,5}{1,3}^{1}}\\
\end{pspicture}\begin{pspicture}(0,0)(0.6,3.8)
\psmatrix[mnode=circle,colsep=0.892] 1 & 2 & 3 & 4 & 5 \\ & 6 & 7 \endpsmatrix
\psset{shortput=nab,arrows=->,labelsep=2pt} {\footnotesize                                   \ncarc[arcangle=10]{1,2}{1,1}^{2}                       \ncline{2,2}{2,3}^{1}
\ncline{1,3}{2,2}_{2}                       \ncline{1,3}{1,4}^{2}                                                                               \ncarc[arcangle=36]{1,5}{1,3}^{1}}\\
\end{pspicture}}
\vspace*{-1em}{\footnotesize\ Forest \#29; weight$\;=\!12$\hspace*{1.9em}\! {\iI}Forest \#30; weight$\;=\!32$\hspace*{1.9em}\!\! {\iI}Forest \#31; weight$\;=\!48$\hspace*{1.9em} {\iI}Forest \#32; weight$\;=\!8$}\\

%\centerline
\noindent
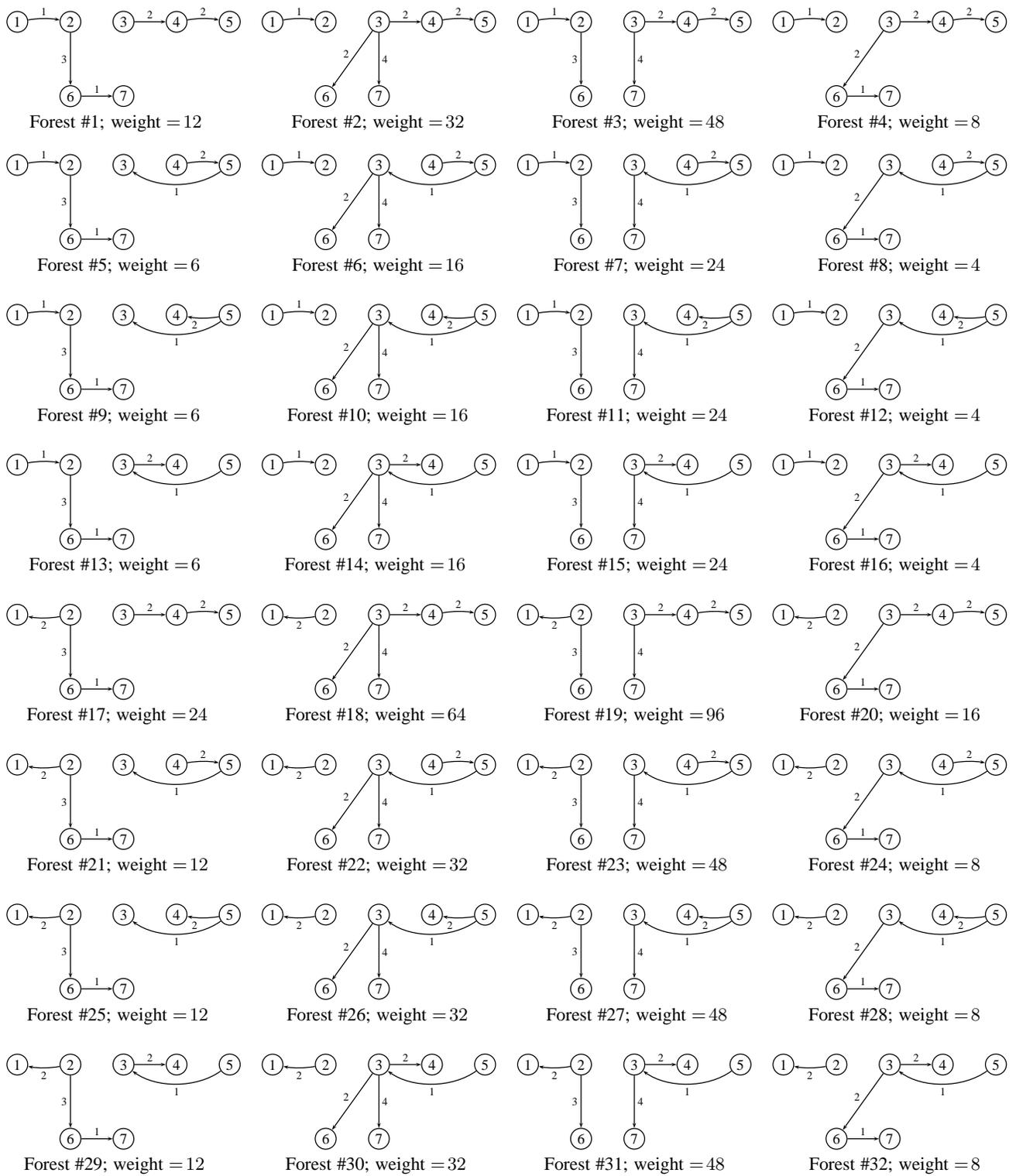
{{\footnotesize Figure 2. The maximum out-forests of the communication digraph in the Example.}}
\vspace*{-2em}
%\caption{\small\label{fig_ex2}The maximum out-forests of the communication digraph in the Example.}

\pagebreak

Using \cite[Proposition\:9]{AgaChe00} the set of all maximum out-forests of $\G$ can be described as follows. 1.~Choose an arbitrary spanning diverging tree in each basic bicomponent of~$\G.$ 2.~Choose any maximum out-forest in the digraph obtained from $\G$ by removing all arcs belonging to the basic bicomponents. Combining the chosen trees and forest gives a maximum out-forest of $\G;$ every desirable forest can be obtained in this way. A~more detailed algorithm for constructing maximum out-forests can be found in \cite[Section\:5]{AgaChe00}.

Let $x(0)=(1, 10, 5, 7, 9, *, *)^T$ (the last two components are ``free'': they correspond to ``nonbasic'' vertices which, by Corollary\:\ref{c_prop}, do not affect the limiting state vector). By Theorem\;\ref{th_main},
\[
\lim_{t\to\infty}x(t)=\vj x(0)=(7,7,7,7,7,7,7)^T,
\]
i.e., asymptotic consensus is achieved. On the other hand, if $x(0)=(0, 6, 3, 9, 10, *, *)^T,$ then
\[
\lim_{t\to\infty}x(t)=\vj x(0)=(4,4,7,7,7,5.2,6.64)^T,
\]
and asymptotic consensus is achieved only within the basic bicomponents having vertex sets $\{1,2\}$ and $\{3,4,5\},$ but not for the whole set of agents.

Thus, a system satisfying \eqref{e_model1}--\eqref{e_model2} has its \emph{domain of convergence to consensus}, that is, the set of initial states $x(0)$ such that the product $\vj x(0)$ gives a vector with all equal components. In \cite{AgaChe11ARCE1}, this domain (obviously, it is a subspace of $\R^n$) is characterized and it is shown that when $x(0)$ does not belong to the domain, then there is still some reasonable ``quasi-consensus''. It can by obtained by first, projecting $x(0)$ onto the domain of convergence and second, applying the coordination protocol \eqref{e_model1}--\eqref{e_model2}. Say, for the initial states of the form $x(0)=(0, 6, 3, 9, 10, *, *)^T$ which were considered above, the value of such a ``quasi-consensus'' is~$5.82$.

%Note that $Z_{10}$ can also be computed using Proposition~1 in~\cite{AgaChe02}.

\section{On communication digraphs of out-forest dimension~1}
\label{s_d=1}

Suppose that the communication digraph $\G$ has a spanning diverging tree or, equivalently, the out-forest dimension of $\G$ is one ($d=1$).
In this (and only this) case, by item\:\ref{i_ra} of Proposition\:\ref{p_eigproj}, $\rank\J=1$ holds, so by item\:\ref{i_1},
\beq{e_J1}
\J=\bm1v_l^T,
\eeq
where $v_l^T$ is any row of~$\J$ and $v_l^T\bm1=1.$ By items\:\ref{i_0} and\;\ref{i_ra}, $v_l^T$ and $\bm1$ span the left and right null spaces of $L,$ respectively. Thus, Theorem\;\ref{th_main} yields the following familiar necessary and sufficient condition of achieving consensus.
\begin{corol}[of Theorem\:\ref{th_main}]\label{c_d1}
If the communication digraph\/ $\G$ of the model \eqref{e_model1}--\eqref{e_model2} has a spanning diverging tree$,$ then for any initial state $x(0),$ $x(t)$ converges to the consensus
\beq{e_Asymp1}%+
\lim_{t\to\infty}x(t)=(v_l^Tx(0))\,\bm1,
\eeq
where $v_l$ is the unique left eigenvector of\/ $L$ associated with\/ $0$ and satisfying $\,v_l^T\bm1=1.$ Conversely$,$ if for each initial state $x(0),$ $x(t)$ tends to a consensus$,$ then $\G$ has a spanning diverging tree.
\end{corol}

For the more restricted case of a strongly connected digraph $\G,$ a representation similar to \eqref{e_Asymp1} was obtained in \cite[Theorem\:3]{Olfati-SaberMurray04}.
In this case, it was shown that $\lim_{t\to\infty}e^{-Lt}=v_rv_l^T,$ where $v_l$ and $v_r$ are, respectively, the left and right eigenvectors of $L$ associated with $0$ and satisfying $v_l^Tv_r=1.$ Before Theorem\:1, the authors of \cite{Olfati-SaberMurray04} mention that $\bm1$ can be substituted for~$v_r.$

Corollary\;\ref{c_d1} coincides with \cite[Theorem\:3.12]{MesbahiEgerstedt10book} (see also \cite[Proposition\:3.11]{MesbahiEgerstedt10book} and Lemma\;1.3 in \cite{RenCao11distributed}).
The case of a communication digraph $\G$ having a spanning diverging tree was recently considered in~\cite{AmelinyGranichiny12} where Lemma~3 presents an analog of \eqref{e_Asymp1}. However, the multiplier $1/\sqrt{n}$ in \cite[Eq.\:(18)]{AmelinyGranichiny12} is not correct due to an invalid step in the proof.

Finally, observe that Theorem\:III.8 in~\cite{TaylorBeardHumpherys11ACC} can also be derived from Theorem\:\ref{th_main}.

\section{A discrete counterpart of the forest consensus theorem}
\label{s_DeGroot}

Consider the discretization of the model\;\eqref{e_modeL}:
\beq{e_discr}%+
\frac{x(t+\tau)-x(t)}{\tau}=-Lx(t)
\eeq
with a small fixed $\tau\in\R.$
Let $y(k)\xz:=\xz x(k\tau),$ $k=0,1\cdc$ be the state vector with the discrete-time dynamics determined by~\eqref{e_discr}.
Then
\beq{e_discry}%-leave
y(k)=(I-\tau L)\,y(k-1),\quad k=1,2,\ldots.
\eeq
Setting
\beq{e_P}%+
P:=I-\tau L
\eeq
and observing \cite[Section\:8]{AgaChe00} that $P$ is row stochastic whenever
\beq{e_tau}%+
0<\tau\le\!\Bigl(\max_i\sum_{j\ne i}a_{ij}\Bigr)^{-1}
\eeq
we obtain De\-Groot's iterative pooling model \cite{DeGroot74}:
\beq{e_DeGroot}
y(k)=P\, y(k-1),\quad k=1,2,\ldots.
\eeq
Matrix \eqref{e_P} is sometimes called the \emph{Perron matrix with parameter $\tau$} of the weighted digraph~$\G.$
Obviously, $P$ is the linear part of the series expansion of $e^{-\tau L}.$

Let us compare the asymptotic properties of the model \eqref{e_modeL} and its discrete analog~\eqref{e_DeGroot}.
From \eqref{e_DeGroot} one has
\beq{e_Pm}%+
y(k)=P^ky(0),\quad k=0,1,\ldots.
\eeq
A necessary and sufficient condition of the convergence of $\{P^k\}$ under \eqref{e_tau} is the aperiodicity of~$P.$ On the other hand, the Ces\`aro (time-average) limit
\beq{e_Pinfry}%+
P^{\infty}:=\lim_{k\to\infty}\frac{1}{k}\sum\limits_{i=1}^k{P^i}
\eeq
exists for any stochastic $P$ and coincides with $\,\lim_{k\to\infty}P^k$ whenever the latter exists. Otherwise, if $P$ is periodic with period $s$, then
$P^{\infty}=s^{-1}\!\left(P^{(1)}+\ldots+P^{(s)}\right),$
where $P^{(1)},\dots,P^{(s)}$ are the limits of the converging subsequences of $\{P^k\}$: $P^{(i)}=\liml_{j\to \infty}P^{js+i},\,$ $i=1\cdc s.$

The discrete-time counterpart of Theorem\;\ref{th_main} is an immediate consequence of well-known results. Yet, for ease of comparison with Theorem\;\ref{th_main}, we represent it in the form of a theorem.

\begin{theorem}
\label{th_main1}
Let sequence $y(k)$ satisfy \eqref{e_DeGroot}$,$ where $P$ is defined by \eqref{e_P}--\eqref{e_tau}. Then
\[%\beq{e_AsympStateD1}%-
\lim_{k\to\infty}\frac{1}{k}\sum\limits_{i=1}^ky(i)=\vj y(0),
\]%\eeq
where $\vj$ is the eigenprojection of\/~$L,$ which coincides with the matrix $\J$ of maximum out-forests of the communication digraph\/ $\G$ corresponding to~$L.$
Moreover$,$ if \eqref{e_tau} is satisfied strictly$,$ then
\beq{e_AsympStateD2}%+
\lim_{k\to\infty}y(k)=\vj y(0).
\eeq
\end{theorem}

\noindent
\textbf{Proof.} %of theorem \ref{th_main1}.}
Meyer \cite{Meyer75} and Rothblum \cite{Rothblum76ai} have shown that $P^{\infty}$ is the eigenprojection of~$I-P$. Hence, by \eqref{e_P} and the definition of eigenprojection$,$ $P^{\infty}=\vj.$ Now applying the Ces\`aro limit to \eqref{e_Pm} and using \eqref{e_Pinfry} and item\;\ref{i_ep} of Proposition\;\ref{p_eigproj} one obtains the first assertion of Theorem\;\ref{th_main1}.

Alternatively, the identity $P^{\infty}=\J$ coincides with the Markov chain tree theorem first proved by Wentzell and Freidlin \cite{WentzellFreidlin70a} and rediscovered
in~\cite{LeightonRivest83,LeightonRivest83a}. This identity provides the first assertion of Theorem\;\ref{th_main1} along the same lines.

Finally, if \eqref{e_tau} is satisfied strictly, then $P$ has a strictly positive diagonal. In this case, by Ger\v{s}gorin's theorem, $P$ has no eigenvalues of modulus $1$ except for~$1.$ Hence, $P$ is not periodic and $\{P^k\}$ converges, which yields~\eqref{e_AsympStateD2}.
\epr

Obviously, the only essential difference between Theorem\;\ref{th_main1} and Theorem\;\ref{th_main} is the use of the Ces\`aro limit in the case of a periodic matrix~$P.$
With a similar ``Ces\`aro'' addendum one can easily formulate a discrete-time counterpart of the Corollary\;\ref{c_d1} of Section\:\ref{s_d=1}.

To compute the matrix $\vj=\J$, one can use items\;\ref{i_li}--\ref{i_01} of Proposition\;\ref{p_eigproj}, constructive characterizations (h), (j) or (l) in \cite[Section\:2]{AgaChe02}, or \cite[Proposition\:2]{CheAga11ARC}.

%\section{Conclusion}
%\label{s_concl}

\bibliographystyle{IEEEtranS}
\bibliography{all2}
\end{document}